\documentclass[preprint,12pt]{elsarticle} 
\usepackage{graphicx} 
\usepackage{amssymb} 
\usepackage{amsmath} 
\usepackage{amsfonts} 
\usepackage{slashed} 
\usepackage[dvipsnames]{xcolor} 
\usepackage{hyperref}
\usepackage{url}
\hypersetup{colorlinks=true,citecolor=blue,urlcolor=blue,linkcolor=blue} 

\newcommand{\be}{\begin{equation}} 
\newcommand{\ee}{\end{equation}}

\def\lamb#1#2{$^{#1}_{\Lambda}${#2}}

\journal{Nuclear Physics A} 

\begin{document} 

\begin{frontmatter}

\title{Remarks on MAMI's recent determination of $B_{\Lambda}$(\lamb{3}{H})} 

\author[a]{Avraham Gal\corref{cor1}} 
\address[a]{Racah Institute of Physics, The Hebrew University, 9190400 
Jerusalem, Israel} 
\cortext[cor1]{Corresponding author: Avraham Gal, avragal@savion.huji.ac.il}  

\begin{abstract} 
A recent report on ${^7{\rm Li}}(e,e'K^+)$ electroproduction runs 
by the A1 collaboration at the Mainz Microtron (MAMI) assigns 
a sharp pion-momentum line at $p_{\pi^-}\approx 113.8\pm 0.1$ MeV/c 
to \lamb{3}{H}$\,\to\pi^-+{^3{\rm He}}$ weak decay, 
resulting in exceptionally large \lamb{3}{H} binding-energy 
$B_{\Lambda}$(\lamb{3}{H})$=0.523\pm 0.013\pm 0.075$~MeV. Here I discuss 
an alternative interpretation of the observed sharp line in terms of 
\lamb{7}{He}$_{\rm g.s.} \to \pi^- + {^7{\rm Li}}(E_{\rm x}=478$~keV) 
weak decay and its implications for $B_{\Lambda}$(\lamb{7}{He}). 
\end{abstract} 


\end{frontmatter}

\section{Introduction}
\label{sec:intro}

\begin{figure}[!h] 
\begin{center} 
\includegraphics[width=0.9\textwidth]{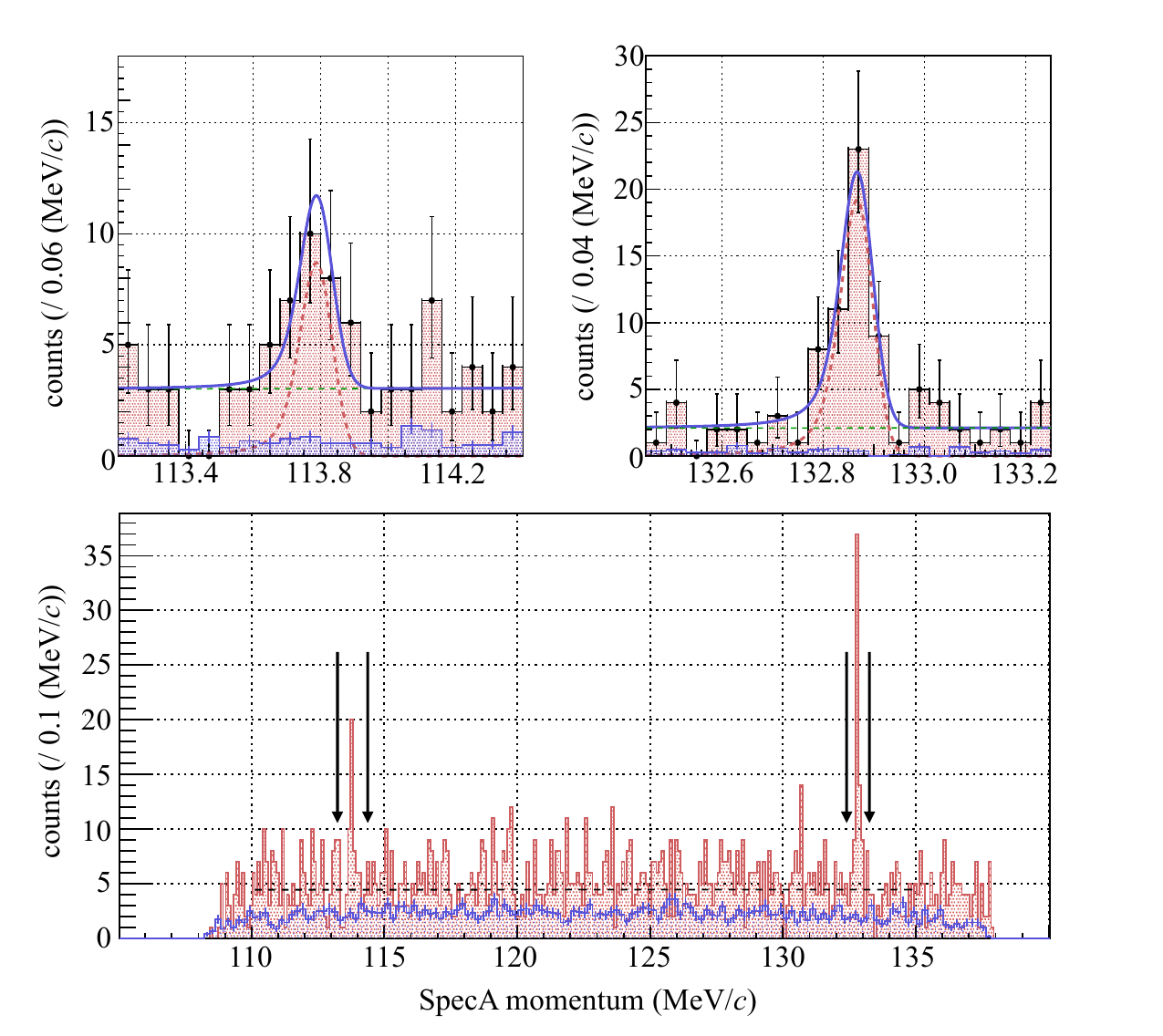}  
\caption{MAMI's decay-pion $p_{\pi^-}$ distribution following $(e,e'K^+)$ 
electroproduction on $^7{\rm Li}$. Lower panel: spectra for true (red) and 
accidental (blue) coincidences. Upper panel: magnified view near $p_{\pi^-}
$=113.8 and 132.9 MeV/c assigned to \lamb{3}{H} and \lamb{4}{H} decays, 
respectively, along with results of unbinned fits. Figure adapted from 
Ref.~\cite{MAMI26}.} 
\label{fig:MAMI26} 
\end{center} 
\end{figure}

Accurate binding energies of the lightest hypernuclei \lamb{3}{H} and 
\lamb{4}{H}-\lamb{4}{He} are essential input for placing EFT constraints 
on $\Lambda N$ and $\Lambda NN$ interactions \cite{CBG18,Le25} that enter 
quantitative discussions of the 'hyperon puzzle'~\cite{Weise20}. In this 
context MAMI's A1 collaboration ${^9{\rm Be}}(e,e'K^+)$ electroproduction 
experiment already provided a most accurate value of $B_{\Lambda}
$(\lamb{4}{H}), the $\Lambda$ binding energy in \lamb{4}{H}, by observing 
a sharp $p_{\pi^-}\approx 132.9$~MeV/c line arising from \lamb{4}{H}$\,\to
\pi^-+{^4}$He weak decay~\cite{MAMI15,MAMI16}. MAMI's very recent 
work~\cite{MAMI26}, using $^7{\rm Li}$ target instead of $^9{\rm Be}$, 
confirms this \lamb{4}{H} sharp line while reporting also another sharp line 
at $p_{\pi^-}=113.789\pm 0.020({\rm stat.})\pm 0.112({\rm syst.})$~MeV/c  
assigned to \lamb{3}{H}$\,\to\pi^-+{^3}$He weak decay, thereby resulting in 
$B_{\Lambda}$(\lamb{3}{H})=0.523$\pm$0.013$\pm$0.075~MeV. This exceptionally 
large \lamb{3}{H} binding-energy is incompatible with all EFT studies in 
the last two decades~\cite{HMN26}. It is more than $4\sigma$ away from 
the weighted average value ${\bar B}_{\Lambda}$(\lamb{3}{H})=0.108$\pm
$0.027~MeV~\cite{KPH26} that {\it preceded} the announcement of MAMI's new 
value. For comparison, this average ${\bar B}_{\Lambda}$ corresponds to 
a weak-decay pion momentum ${\bar p}_{\pi^-}=114.40\pm 0.04$~MeV/c, quite 
distinct from MAMI's sharp $p_{\pi^-}\approx 113.8\pm 0.1$~MeV/c line. 
MAMI's decay-pion $p_{\pi^-}$ distribution, using a $^7{\rm Li}$ target, 
is shown in Fig.~\ref{fig:MAMI26}. 

\begin{figure}[!ht] 
\begin{center} 
\includegraphics[angle=270,width=0.9\textwidth]{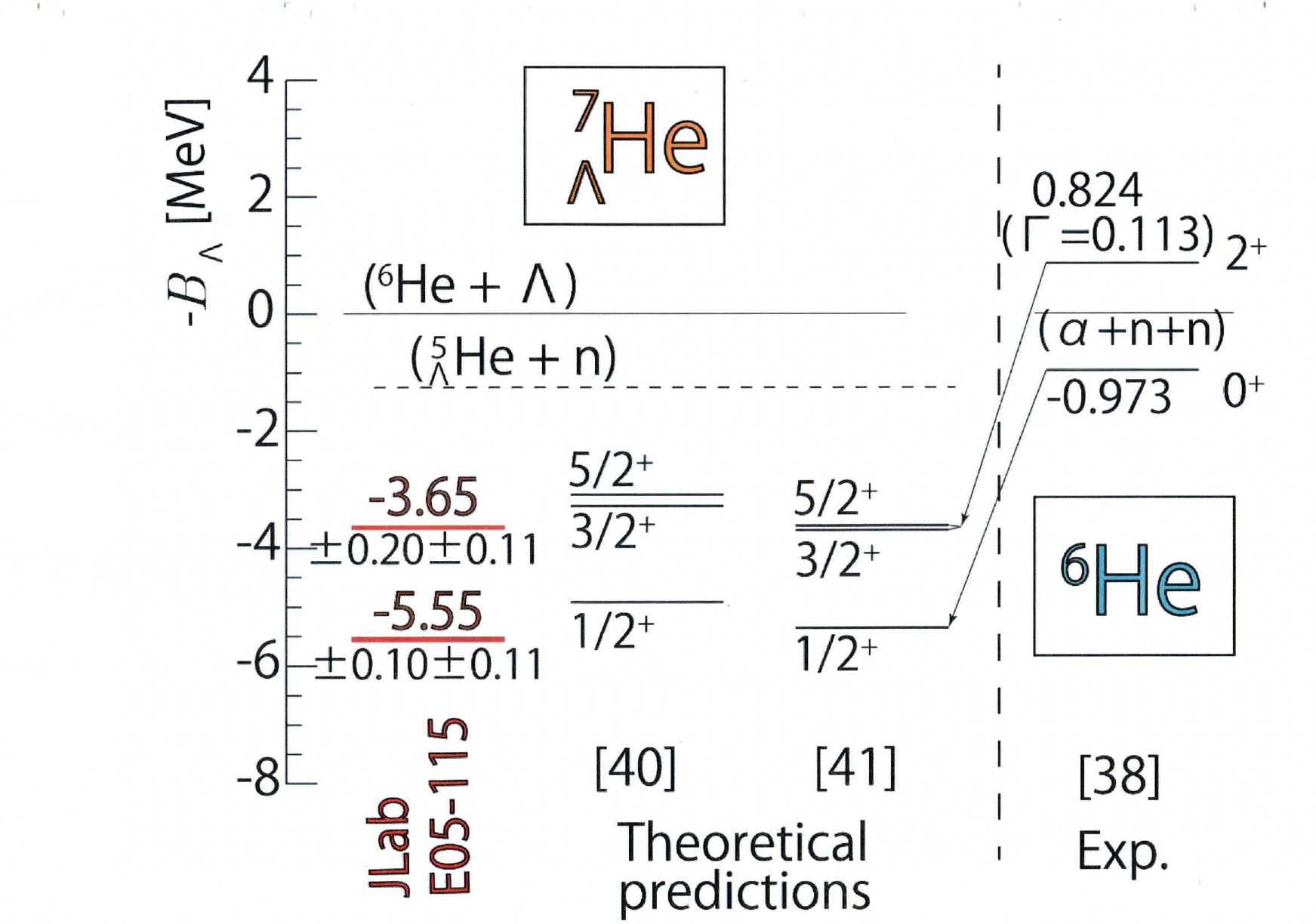} 
\caption{Particle-stable \lamb{7}{He} levels obtained in JLab Hall C E05-115 
$(e,e'K^+)$ electroproduction experiment on $^7$Li target, along with two 
calculated spectra. The $^6$He nuclear-core spectrum is also shown. Figure 
adapted from Ref.~\cite{Gogami16}.}   
\label{fig:L7He} 
\end{center} 
\end{figure}

The ${^7{\rm Li}}(e,e'K^+)$ large momentum-transfer electroproduction reaction 
converts a bound proton to a $\Lambda$ hyperon within a broad spectrum of 
bound and continuum \lamb{7}{He} states. Bound-state levels of \lamb{7}{He} 
established in JLab experiments are shown in Fig.~\ref{fig:L7He}. Assigning 
the sharp line observed at $p_{\pi^-}\approx 113.8\pm 0.1$~MeV/c~\cite{MAMI26} 
to \lamb{7}{He}$_{\rm g.s.}\to\pi^-+{^7}{\rm Li}_{\rm g.s.}$ g.s.-to-g.s. 
weak decay leads to a value of $B_{\Lambda}$(\lamb{7}{He})=6.32$\pm$0.09~MeV, 
much higher than the two JLab values 5.68$\pm$0.25~MeV~\cite{Nakamura13} and 
5.55$\pm$0.15~MeV~\cite{Gogami16}. Notably, these JLab values correspond to 
a weak-decay pion momentum $p_{\pi^-}\approx 115$~MeV/c where no pronounced 
signal is observed; see lower panel of Fig.~\ref{fig:MAMI26} here and also 
Fig.~1 in Ref.~\cite{MAMI26_com}. 

Realizing that the $p_{\pi^-}\approx 113.8$~MeV/c sharp line does not 
correspond to \lamb{7}{He}$_{\rm g.s.}\to\pi^-+{^7}{\rm Li}_{\rm g.s.}$ weak 
decay, we explore below two-body weak decays involving {\it excited} states 
of \lamb{7}{He} and/or \lamb{7}{Li}. Indeed, considering theoretical model 
uncertainties in the actual value of $B_{\Lambda}$(\lamb{7}{He}), 
we do find that one such weak-decay is compatible with MAMI's~\cite{MAMI26} 
$p_{\pi^-}\approx 113.8$~MeV/c sharp line. 

As for \lamb{7}{He} continuum states, we focus on two-body breakup channels 
through which lighter hypernuclei are most likely formed. This is suggested by 
inspecting a statistical decay model yields calculated for electroproduction 
on a $^9$Be target~\cite{MAMI12,MAMI13}. There are just two such 
breakup channels for \lamb{7}{He} continuum states: a low-lying \lamb{6}{He}+n 
channel with threshold at $-$2.4 MeV and a \lamb{4}{H}+$^3$H channel with 
threshold at 10.1 MeV, both with respect to the ${^6}$He+$\Lambda$ threshold. 
Barring \lamb{6}{He}$\,\to\pi^-+{^6}$Li decay, with $p_{\pi^-}\approx 108.5$ 
MeV/c almost outside the $p_{\pi^-}$ momentum range in Fig.~\ref{fig:MAMI26}, 
one is left with only \lamb{4}{H} as a candidate from \lamb{7}{He} continuum 
states for a $\pi^-$ decay signal. For \lamb{3}{H}, its formation requires 
a three-body breakup channel, \lamb{3}{H}+$^3$H+n, with a much higher 
threshold at 21 MeV. We note that \lamb{3}{H} production is more likely 
in electroproduction on $^9$Be target by forming a \lamb{9}{Li} spectrum of 
bound and continuum states with several two-body breakup channels. Two such 
channels, \lamb{7}{He}+d and $^6$He+\lamb{3}{H}, arise from the 9.784~MeV 
$^6$He+d threshold in $^8$Li. However, applying a statistical decay model to 
the breakup of \lamb{9}{Li} continuum states, it appears that production of 
\lamb{7}{He} marked by a weak-decay pion momentum $p_{\pi^-}\approx 115$~MeV/c 
is an order of magnitude stronger than production of \lamb{3}{H} marked by 
the nearby pion momentum $p_{\pi^-}\approx 114$~MeV/c~\cite{MAMI12,MAMI13}. 
The suppressed yield of \lamb{3}{H} is due apparently to its halo nature. 
Analogous \lamb{7}{He} and \lamb{3}{H} yield calculations for MAMI's $^7$Li 
target are not available. This concerns mostly the production of \lamb{3}{H} 
since we know from JLab electroproduction experiments on $^7$Li that 
\lamb{7}{He} bound states allow observation~\cite{Nakamura13,Gogami16}. 

Having argued that a sharp $p_{\pi^-}$ line in ${^7{\rm Li}}(e,e'K^+)$ 
electroproduction is expected from \lamb{4}{H}$\,\to\pi^- + {^4}$He two-body 
weak decay, it is clear that the sharp $p_{\pi^-}\approx 132.9$~MeV/c line in 
Fig.~\ref{fig:MAMI26} indeed corresponds to \lamb{4}{H} weak decay, as already 
established in MAMI's runs on $^9$Be target~\cite{MAMI15,MAMI16}. No other 
$p_{\pi^-}$ weak-decay line candidates are available around this $p_{\pi^-}$ 
value.

\section{\lamb{7}{He}$\to \pi^- + {^7}{\rm Li}$ weak decays}
\label{sec:weak} 

In this section we consider {\it all} possible \lamb{7}{He}$\to\pi^-+ {^7}$Li 
weak decays. For \lamb{7}{He}, apart from its (1/2)$^+_{\rm g.s.}$ level we 
also consider a (5/2)$^+$ excited level at excitation energy $E_{\rm x}\approx 
1.9$~MeV, see Fig.~\ref{fig:L7He}. This (5/2)$^+$ level is degenerate, within 
experimental resolution, with a (3/2)$^+$ level~\cite{Gogami16}. Both are 
obtained to a good approximation by coupling a $1s_{\Lambda}$ to the 
($2^+,E_{\rm x}=1.797$~MeV) first excited $^6$He level. The (5/2)$^+$ level is 
likely to be isomeric, as first argued by Pniewski and Danysz~\cite{PD62}. 
Consisting of three neutral particles outside of a strongly bound $^4$He core, 
the \lamb{7}{He} $(5/2)^+ \to (1/2)^+_{\rm g.s.}$ E2 electromagnetic (e.m.) 
transition lifetime is longer than the $\Lambda$ weak-decay lifetime 
$\tau_{\Lambda}=2.63\times 10^{-10}$~sec~\cite{Elt62}. Its M1 e.m. transition 
to the (3/2)$^+$ level, provided (5/2)$^+$ lies above (3/2)$^+$, is also 
longer than $\tau_{\Lambda}$ owing to the small (3/2)$^+$-(5/2)$^+$ doublet 
separation of order 0.1~MeV predicted by using Millener's best fit of 
$\Lambda N$ spin dependent parameters to observed $\gamma$-ray transition 
energies in the $p$ shell~\cite{DJM08}. This scenario, involving some caveats, 
was discussed in several Dalitz-Gal papers~\cite{DG67,DG78a,DG78b}. 

For ${^7}$Li, apart from its (3/2)$^-_{\rm g.s.}$ level we also consider 
the particle-stable $(1/2)^-$ first excited level at $E_{\rm x}=0.478$~MeV. 
These two lowest ${^7}$Li bound states form to a good approximation a ${^2}P$ 
LS doublet. The next LS doublet, ${^2}F$, consists of a (7/2)$^-$ level at 
$E_{\rm x}$=4.652~MeV and a (5/2)$^-$ level at $E_{\rm x}$=6.604~MeV. Both are 
particle-unstable, but only the (7/2)$^-$ level is perhaps sufficiently narrow 
($\Gamma$=69~keV~\cite{TUNL}) to be considered here. 

\begin{table}[!htb] 
\centering 
\caption{Pion momentum lines $p_{\pi^-}$ expected in \lamb{7}{He}$\to\pi^- + 
{^7}{\rm Li}$ weak decays, assuming: (i) $B_{\Lambda}$(\lamb{7}{He})=5.55$\pm
$0.15~MeV from a recent JLab ${^7{\rm Li}}(e,e'K^+)$\lamb{7}{He} 
electroproduction run~\cite{Gogami16}; (ii) $B_{\Lambda}$(\lamb{7}{He})=5.84$
\pm$0.07~MeV, see text.} 
\begin{tabular}{cccc} 
\hline 
\lamb{7}{He}: $J_i^{\pi}$, $E_{\rm x}$ (MeV) & $^7$Li: $J_f^{\pi}$, 
$E_{\rm x}$ (MeV) & $p^{(i)}_{\pi^-}$ (MeV/c) & $p^{(ii)}_{\pi^-}$ (MeV/c) \\ 
\hline 
(1/2)$^+_{\rm g.s.}$ & (3/2)$^-_{\rm g.s.}$ & 114.98$\pm$0.24 & 
114.53$\pm$0.11 \\ 
(1/2)$^+_{\rm g.s.}$ & (1/2)$^-$, 0.478 & 114.24$\pm$0.24 & 
113.79$\pm$0.11 \\ 
(5/2)$^+$, 1.90$\pm$0.23 & (3/2)$^-_{\rm g.s.}$ & 117.86$\pm$0.35 & 
117.42$\pm$0.35 \\ 
(5/2)$^+$, 1.90$\pm$0.23 & (7/2)$^-$, 4.652 & 110.71$\pm$0.36 & 
110.26$\pm$0.36 \\ 
\hline 
\end{tabular} 
\label{tab:p(pi)} 
\end{table} 

Several pion-momentum $p_{\pi^-}$ values expected in \lamb{7}{He}$\to\pi^- + 
{^7}{\rm Li}$ weak decays of both \lamb{7}{He} (1/2)$^+_{\rm g.s.}$ and 
(5/2$^+,E_{\rm x}=1.90\pm 0.23$~MeV) levels, assuming the latter is isomeric, 
are listed in Table~\ref{tab:p(pi)}. For ${^7}{\rm Li}$ final states we 
consider the three lowest levels, including the (7/2$^-,E_{\rm x}=4.652$~MeV) 
level mentioned above. Input hypernuclear masses were evaluated using binding 
energies $B_{\Lambda}$(\lamb{7}{He}), including uncertainties, from the most 
recent JLab ${^7{\rm Li}}(e,e'K^+)$\lamb{7}{He} electroproduction experiment 
\cite{Gogami16}. The resulting $p_{\pi^-}$ values are listed in the 
third column. One such value, $p_{\pi^-}$=114.24$\pm$0.24~MeV/c for 
\lamb{7}{He}(1/2$^+_{\rm g.s.})\to\pi^-+{^7}$Li(1/2$^-,E_{\rm x}=0.478$~MeV), 
is quite close to the $p_{\pi^-}$=113.79$\pm$0.11~MeV/c sharp line in 
Fig.~\ref{fig:MAMI26} which would require a value of $B_{\Lambda}
$(\lamb{7}{He})=5.84$\pm$0.07~MeV, almost 2$\sigma$ away from JLab's 
5.55$\pm$0.15~MeV~\cite{Gogami16}. The $p_{\pi^-}$ values corresponding 
to $B_{\Lambda}$(\lamb{7}{He})=5.84$\pm$0.07~MeV are listed in the fourth 
column. We note that MAMI's measured pion spectra show no signal whatsoever 
for $p_{\pi^-}$ values in the interval 114 to 115 MeV/c~\cite{MAMI26_com} 
where according to Table~\ref{tab:p(pi)} the \lamb{7}{He}(1/2$^+_{\rm g.s.}$) 
pionic weak decay to $^7$Li(3/2$^-_{\rm g.s.}$) is expected with 
(following Ref.~\cite{Gal09}) a rate about twice stronger than to 
the (1/2$^-,E_{\rm x}=0.478$~MeV) level. The absence of this 
\lamb{7}{He}(1/2$^+_{\rm g.s.})\to\pi^-+^7$Li(3/2$^-_{\rm g.s.}$) line in 
the reported spectrum, posing a problem for our \lamb{7}{He} interpretation, 
could be further studied by increasing gradually the 2~ns limit of the 
lifetime gate imposed in the MAMI experiment~\cite{MAMI26}. 

As for $p_{\pi^-}$ lines from weak decay of the \lamb{7}{He}(5/2$^+,E_{\rm x}
=1.90$~MeV) level, we considered decays to the (3/2)$^-_{\rm g.s.}$ and 
(7/2$^-,E_{\rm x}=4.652$~MeV) levels in $^7$Li. These two decays, as well as 
the two decays from \lamb{7}{He}(1/2$^+_{\rm g.s.}$) considered above, proceed 
dominantly through the leading $\Delta L=1 (1s_{\Lambda}\to 1p_{\Lambda})$, 
$\Delta S=0$ transition. In contrast, the \lamb{7}{He}(5/2$^+,E_{\rm x}=1.90
$~MeV)$\to \pi^-+{^7}$Li(1/2$^-,E_{\rm x}=0.478$~MeV) weak decay requires a 
subleading $\Delta S=0$ transition, resulting in a suppressed decay rate with 
lifetime perhaps even longer than the 2~ns lifetime gate. 

MAMI's decay-pion momentum distribution shown in Fig.~\ref{fig:MAMI26} offers 
no visible peaks for $\pi^-$ weak decay from the (5/2)$^+$ level. Of the two 
listed candidates, the calculated weak-decay rate to $^7$Li(7/2$^-,E_{\rm x}=
4.652$~MeV) is by far the stronger one, close to that calculated for the 
\lamb{7}{He}(1/2$^+_{\rm g.s.})\to\pi^-+{^7}$Li(1/2$^-,E_{\rm x}=0.478$~MeV) 
decay rate, as may be verified by following Ref.~\cite{Gal09}. The absence of 
a candidate $p_{\pi^-}$ line in Fig.~\ref{fig:MAMI26} may be related to the 
non-negligible 69~keV width of the $^7$Li(7/2$^-$) final state. Recall also 
that the (3/2$^+$,5/2$^+$) formation differential cross section in 
${^7{\rm Li}}(e,e'K^+)$\lamb{7}{He} is $\sim$0.6 of that for the 
(1/2)$^+_{\rm g.s.}$~\cite{Gogami16}, with the (5/2)$^+$ formation rate 
suppressed further by another 0.6 factor, proportional to its $(2J+1)$ 
relative weight, suggesting thereby that its formation rate is only 0.36 
of the formation rate of the (1/2)$^+_{\rm g.s.}$ level.

\section{\lamb{7}{He} binding energy} 
\label{sec:L7He} 

We argued above that \lamb{7}{He} is likely to have been formed in MAMI's 
${^7{\rm Li}}(e,e'K^+)$ experiment~\cite{MAMI26} rather than \lamb{3}{H}. 
Relating the observed sharp $p_{\pi^-}\approx 113.8\pm 0.1$~Mev/c line to 
\lamb{7}{He}$_{\rm g.s.}\to\pi^-+{^7}$Li(1/2$^-,E_{\rm x}=0.478$~MeV) 
weak decay requires a value of $B_{\Lambda}$(\lamb{7}{He})=5.84$\pm$0.07~MeV. 
In this section we discuss constraints imposed on $B_{\Lambda}$(\lamb{7}{He}) 
by the other A=7 hypernuclei. Thus, recall that \lamb{7}{He}$_{\rm g.s.}$, 
\lamb{7}{Li*}, and \lamb{7}{Be}$_{\rm g.s.}$, where \lamb{7}{Li*} is 
\lamb{7}{Li}($E_{\rm x}$=3.877~MeV)~\cite{E419}, all with $J^{\pi}$=1/2$^+$, 
form a $T=1$ isospin triplet with $T_z=-1,0,+1$, respectively. \lamb{7}{Li*} 
is well approximated by coupling a $1s_{\Lambda}$ to the first $^6$Li $T=1$ 
level ($0^+,E_{\rm x}=3.563$~MeV). $B_{\Lambda}$(\lamb{7}{Li*}) is then 
obtained by subtracting 3.877$-$3.563=0.314~MeV from $B_{\Lambda}
$(\lamb{7}{Li})=5.58$\pm$0.05~MeV (emulsion value~\cite{Davis05}). This 
results in $B_{\Lambda}$(\lamb{7}{Li*})=5.27$\pm$0.05~MeV. Assuming isospin 
breaking linear in $T_z$ and using $B_{\Lambda}$(\lamb{7}{Be})=5.16$\pm
$0.09~MeV (emulsion value~\cite{Davis05}, there are none else), we get 
$B_{\Lambda}$(\lamb{7}{He})=5.38$\pm$0.10~MeV. This value, listed in the 
first row of Table~\ref{tab:7Li*}, is about $1\sigma$ {\it below} JLab's 
most recent value 5.55$\pm$0.15~MeV~\cite{Gogami16}, but remarkably close 
to $B_{\Lambda}$(\lamb{7}{He})=5.36~MeV obtained in a $\Lambda$+n+n+$^4$He 
four-body calculation by Hiyama et al.~\cite{Hiyama15}. 

\begin{table}[!htb] 
\centering 
\caption{$B_{\Lambda}$(\lamb{7}{He}) values derived by using two different 
$B_{\Lambda}$(\lamb{7}{Li}) input values: from emulsion~\cite{Davis05} (1st 
row) and from a KEK-SKS ($\pi^+,K^+$) spectrum~\cite{GHM16,BBF17} (2nd row); 
see text. Binding energies $B_{\Lambda}$ are given in MeV.} 
\begin{tabular}{cccc} 
\hline 
$B_{\Lambda}$(\lamb{7}{Li}) & $B_{\Lambda}$(\lamb{7}{Be}) & 
$B_{\Lambda}$(\lamb{7}{Li*}) & $B_{\Lambda}$(\lamb{7}{He}) \\ 
\hline 
5.58$\pm$0.05 & 5.16$\pm$0.09 & 5.27$\pm$0.05 & 5.38$\pm$0.10 \\ 
5.82$\pm$0.11 & 5.16$\pm$0.09 & 5.51$\pm$0.11 & 5.86$\pm$0.14 \\ 
\hline 
\end{tabular} 
\label{tab:7Li*} 
\end{table} 

However, the emulsion datum $B_{\Lambda}$(\lamb{7}{Li})=5.58$\pm$0.05~MeV 
used above is in conflict with the KEK-SKS ($\pi^+,K^+$) revised value 
5.82$\pm$0.11~MeV \cite{GHM16,BBF17} and with the DA$\Phi$NE-FINUDA 
($K^-_{\rm stop},\pi^-$) value 5.85$\pm$0.16~MeV \cite{FINUDA09}. Using the 
former value, one obtains $B_{\Lambda}$(\lamb{7}{Li*})=5.51$\pm$0.11~MeV 
which together with  $B_{\Lambda}$(\lamb{7}{Be})=5.16$\pm$0.09~MeV suggest  
$B_{\Lambda}$(\lamb{7}{He})=5.86$\pm$0.14~MeV, as listed in the second 
row of Table~\ref{tab:7Li*}. This value is in perfect agreement with 
$B_{\Lambda}$(\lamb{7}{He})=5.84$\pm$0.07~MeV that according to the 
discussion in Sect.~\ref{sec:weak} is associated with MAMI's observed 
sharp $p_{\pi^-}\approx 113.8\pm 0.1$~MeV line. We note that both 
$B_{\Lambda}$(\lamb{7}{He}) values listed in the last column of 
Table~\ref{tab:7Li*} are compatible, within uncertainties, with the 
NLO19+YNN EFT calculated value 5.64$\pm$0.27~MeV~\cite{LHMN23}. 

The discussion above assumes a spin-independent ($T=1,A=7$) hypernuclear mass 
formula linear in the isospin projection $T_z$. This provides a good 
approximation in cases where the corresponding $T=1$ nuclear-core states are 
spin singlet, as is the case for the dominantly $^1S_0$ ($T=1,A=6$) nuclear 
cores ($^6$He,$^6$Li,$^6$Be). In the general case isospin breaking introduces 
additional terms, such as ${\vec{\sigma}}_N\cdot{\vec{\sigma}}_{\Lambda}\,
\tau_{Nz}$, requiring a more specialized discussion.

\section{Summary} 
\label{sec:sum} 

In conclusion, we have presented arguments for associating \lamb{7}{He} with 
the sharp $p_{\pi^-}\approx\,$113.8~MeV/c line reported by MAMI in their 
recent ${^7{\rm Li}}(e,e'K^+)$ electroproduction experiment~\cite{MAMI26}. 
This sharp line could have arisen from the \lamb{7}{He}(1/2$^+_{\rm g.s.})
\to\pi^-+{^7}$Li(1/2$^-,E_{\rm x}=0.478$~MeV) weak decay, provided the binding 
energy of \lamb{7}{He} is 5.84$\pm$0.07~MeV. We have shown how $B_{\Lambda}
$(\lamb{7}{He}) is related to $B_{\Lambda}$(\lamb{7}{Be}) and to $B_{\Lambda}
$(\lamb{7}{Li}), suggesting at present two different values for $B_{\Lambda}
$(\lamb{7}{He}), one of which is consistent with 5.84$\pm$0.07~MeV. 
However, so long as the absence of 
a \lamb{7}{He}(1/2$^+_{\rm g.s.})\to\pi^-+{^7}$Li(3/2$^-_{\rm g.s.}$) 
weak-decay $p_{\pi^-}\approx\,$114.5~MeV/c line is not understood, 
our \lamb{7}{He} interpretation of MAMI's experiment cannot be considered 
conclusive. 

\section*{Acknowledgments}

Many thanks are due to Josef Pochodzalla for useful discussions of MAMI's 
recent findings as well as to Ryoko Kino, first author of Ref.~\cite{MAMI26}, 
who provided me with Fig.~\ref{fig:MAMI26}, and to Toshi Gogami, first author 
of Ref.~\cite{Gogami16} who provided me with Fig.~\ref{fig:L7He}.

\end{document}